\documentstyle[epsfig]{mn}
\begin{document}

\def\ltsima{$\; \buildrel < \over \sim \;$}
\def\simlt{\lower.5ex\hbox{\ltsima}}
\def\gtsima{$\; \buildrel > \over \sim \;$}
\def\simgt{\lower.5ex\hbox{\gtsima}}
\def\ls{{_<\atop^{\sim}}}
\def\lax{{_<\atop^{\sim}}}
\def\gs{{_>\atop^{\sim}}}
\def\gax{{_>\atop^{\sim}}}

\title[ASCA observations of two steep soft X-ray quasars]
{ASCA observations of two steep soft X-ray quasars}

\author[F. Fiore et al.]{F. Fiore$^{1,2,3}$, G. Matt$^4$, M. Cappi$^{5,6}$, 
M. Elvis$^3$, K. M. Leighly$^{6,7}$, F. Nicastro$^{1,8}$, \\ ~ \\
{\LARGE L. Piro$^8$, A. Siemiginowska$^3$, B. J. Wilkes$^3$ } \\ ~ \\
$^1$ Osservatorio Astronomico di Roma, Via dell'Osservatorio,
I--00044 Monteporzio Catone, Italy\\
$^2$ SAX Science Data Center, Via Corcolle 19, I--00131 Roma, Italy\\
$^3$ Harvard-Smithsonian Center of Astrophysics, 60 Garden Street, 
Cambridge MA 02138 USA\\
$^4$ Dipartimento di Fisica, Universit\`a degli Studi ``Roma Tre",
Via della Vasca Navale 84, I--00146 Roma, Italy \\
$^5$ Istituto per le Tecnologie e Studio Radiazioni
Extraterrestri (ITESRE), CNR, via Gobetti 101, I--40129 Bologna, Italy \\
$^6$  Cosmic Radiation Laboratory, RIKEN, Hirosawa 2-1,
Wako-shi, Saitama, 351, Japan \\
$^7$ Columbia Astrophysics Laboratory, Columbia
University, 538 West 120th Street, New York, NY 10027, USA\\
$^8$ Istituto di Astrofisica Spaziale (IAS), CNR, Via E. Fermi 21,
I00044 Frascati, Italy\\
}
\maketitle
\begin{abstract}

Steep soft X-ray (0.1-2 keV) quasars share several 
unusual properties: narrow Balmer lines, strong FeII emission, large
and fast X-ray variability, rather steep 2-10 keV spectrum. 
These intriguing objects have been suggested to be the analogs
of Galactic black hole candidates in the high, soft state. We
present here results from ASCA observations for two of these quasars:
NAB0205+024 and PG1244+026.

Both objects show similar variations (factor of $\sim2$ in 10 ks),
despite a factor of about ten difference in the 0.5-10 keV 
luminosity ($7.3\times 10^{43}$ erg s$^{-1}$ for PG1244+026 and
$6.4\times10^{44}$ erg s$^{-1}$ for NAB0205+024, assuming isotropic
emission, $H_0$ = 50.0 and $q_0$ = 0.0).

The X-ray continuum of the two quasars flattens by 0.5-1 going from
the 0.1-2 keV band toward higher energies, strengthening recent
results on another half dozen steep soft X-ray AGN.

PG1244+026 shows a significant feature in the `1 keV' region, which
can be described by either as a broad emission line centered
at 0.95 keV (quasar frame) or as edge or line absorption at 1.17 (1.22)
keV.  The line emission could be due to reflection from an highly
ionized accretion disk, in line with the view that steep soft X-ray
quasars are emitting close to the Eddington luminosity.  Photoelectric
edge absorption or resonant line absorption could be produced
by gas outflowing at a large velocity (0.3-0.6 c).

\end{abstract}


\section{Introduction}

The ROSAT PSPC has found a large spread in the energy spectral indices of
low-z quasars\footnote{We use ``quasars'' to describe broad line emission 
objects, regardless of luminosity.} : $0.5<\alpha_{0.1-2keV}<3.5$. In about
10$\%$ of cases $\alpha_{0.1-2keV}\gs 2$ (e.g. Laor et al. 1994, 1997,
Walter \& Fink 1993, Fiore et al. 1994). The large spread in
$\alpha_{0.1-2keV}$ favoured the discovery of its correlation with 
other properties. 
In fact, the steep soft X-ray quasars have then been realized to
share a cluster of unusual properties:

\begin{itemize}

\item 
narrow Balmer lines \footnote{the permitted lines have
FWHM$\ls$2000~km~s$^{-1}$, yet still are clearly broader
than the forbidden lines.} ( Laor et al
1994, 1997, Boller et al. 1995);

\item 
strong FeII emission (Laor et al 1994, 1997, Lawrence et al. 1997)

\item Rapid, large amplitude variability (factor of 2-50 on
timescales from minutes to months, Boller et al., 1995,
Brandt et al., 1995, Otani 1995, Boller et al. 1997)

\item Somewhat steep hard X-ray spectra (2$>\alpha_{2-10keV} >0.6$, 
Pounds et al. 1995, Brandt et al., 1997);

\end{itemize}

Pounds et al. (1995), suggest the latter to be a close physical
analogy with the X-ray power-law produced by Comptonization in a hot
accretion disk corona in Galactic black hole candidates (BHC) in their
`soft-high' state. This is not the only analogy between BHC and steep
X-ray spectrum quasars.  Laor et al. (1994, 1997) explained the
correlation with H$\beta$ FWHM as due to the larger size of a
virialized broad emission line region for an AGN in a high $L/L_{Edd}$
state.  Ebisawa (1991) found that while the soft component of 6 BHC
observed by Ginga is roughly stable on time scales of 1 day or less,
the hard component exhibits large variations down to msec time scales.
These timescales translates to $10^4$ years and 0.1 day for quasars,
if they scale with the mass of the compact object.  The soft component
of BHC extends up to $\sim 10$ keV in BHC in `soft-high' states, and
it is often associated with optically thick emission from an accretion
disk. If this is the case, the temperature should scale with the mass
of the compact object as $M_{BH}^{-1/4}$, and the above energy
translates to 0.1-0.4 keV for quasars.  The rapid large amplitude
variability shown by a few narrow line Seyfert 1 galaxies (NLSy1) at
about 1 keV on timescales of hours to days (Otani 1995, Brandt et
al. 1995, Boller et al. 1997) can then be analogous to the above BHC
hard component flickering.

A steep X-ray spectrum quasar with 10-100 times the luminosity
of NLSy1s, should be larger and so should vary no more rapidly
than several days. Instead Fiore et al. (1998a) find that steep X-ray
spectrum PG quasars commonly vary by a factor 2 in 1~day. Variability
seems therefore correlated with X-ray spectral slope and Balmer line
width (and therefore possibly with the accretion rate) rather than
with the luminosity.

Evidence for spectral features in the `1 keV' region in many steep
soft X-ray quasars is building up (Turner et al., 1991, Brandt et al.,
1994, Otani et al., 1995, Comastri et al., 1995, Leighly et al.,
1997, 1998a,b).  Instead, `normal' Seyfert 1 galaxies (having broad
Balmer lines and flatter soft X-ray spectra) usually have their
strongest absorption features at lower energies (in the 0.6-0.9 keV
`oxygen' band).  An intriguing possibility is that the appearance of
these features at different energies also depend on $L/L_{\rm Edd}$.

Detailed high energy X-ray spectra of luminous quasars with steep soft
X-ray spectra are essential to understand the `narrow-broad line'
phenomenon in AGN, in particular whether the peculiar X-ray properties
depend on optical luminosity, optical-to-X-ray ratio ($\alpha_{OX}$),
or on their Eddington ratio.  To this end we selected two bright
quasars with $\alpha_{0.1-2keV}>$2.0 (Fiore et al., 1994) at the extreme
values of optical luminosity, both with low Galactic $N_H$ (Table 1) of
$1.9\times 10^{20}$ cm$^{-2}$ for PG~1244+026, and of 
$3.0\times 10^{20}$ cm$^{-2}$ for NAB0205+024, Elvis et al., 1989)
and observed them with ASCA. We report the results in this paper.

\section{Observation and data reduction}

Table 1 gives the redshift, M$_V$, the 0.2--2 keV 
luminosity, $\alpha_{OX}$, the average PSPC count rate and spectral index
and the Galactic $N_H$ for the two quasars.
Table 2 gives the ASCA observation log, the SIS and GIS 
exposure times and count rates.
 
\begin{table*}
\centering
\caption{Steep soft X-ray quasars}
\begin{tabular}{lccccccc}
\hline \hline
name & z & M$_V^a$ & $L_{0.2-2keV}$ & $\alpha_{OX}$ & PSPC & 
$\alpha_{0.1-2keV}$ & $N_{H(Gal)}$ \\ 
     &   &       & $10^{45}$ erg s$^{-1}$ &      & cts/s & 
                    & $10^{20}$ cm$^{-2}$ \\
\hline
PG1244+026      & 0.048 & -21.1  & 0.14 & 1.4 & 1.0  & $2.3\pm0.1$ & 
1.9$^b$ \\
NAB0205+024     & 0.155 & -25.0  & 1.8  & 1.6 & 0.7  & $2.3\pm0.1$ & 
3.0$^b$ \\
\hline
\end{tabular}

$^a$ $H_0=50,~q_0=0$; $^b$ Elvis et al. 1989

\end{table*}

\begin{table*}
\centering
\caption{ASCA observations}
\begin{tabular}{lccccc}
\hline \hline
name & Dates & Exposure SIS-GIS & 2 SIS count rate & 2 GIS count rate   \\ 
     &       &   ks             & cts/s            &   cts/s         \\
\hline
PG1244+026  & 1-3 Jul 1996 & 37-39 & 0.442$+/-$0.004 & 0.241$+/-$0.004 \\
NAB0205+024 & 18-20 Jan 1996 & 50-54 & 0.150$+/-$0.001 & 0.153$+/-$0.002 \\
\hline
\end{tabular}
\end{table*}

Both observations were performed in two CCD mode with the source at
the `1CCD mode' position. Data reduction was performed using {\sc
ftools 3.6}.  We used "bright" mode SIS data, combining LOW, MEDIUM
and HIGH bit rates.  Conservative cleaning criteria were applied
(minimum Earth occultation = 7 degrees, minimum magnetic rigidity = 6
GeV/c, minimum bright Earth angle = 20 degrees, and excluding data
collected in the first 32 seconds after the satellite passage in the
SAA and through the day-night terminator). Counts, light curves and
spectra from the two quasars were accumulated in circular regions of 3
and 4 arcmin radius for SIS and GIS respectively.

We are interested in the high energy spectrum of these sources and
since they are rather faint, and possibly very steep, background
subtraction plays a crucial role.  Background counts were accumulated
from regions surrounding the sources and compared with counts
accumulated from the same regions from `blanksky' observations. The
`local' and `blanksky' background counts were always within 10 \% of
each other for the four ASCA instruments.  To obtain the best possible
signal to noise in the background subtracted spectra we therefore used
the `blanksky' background in our spectral analysis.  We extracted
background spectra from `blanksky' event files using the same regions
as for source extraction.  The count rates of the two sources become
that of the background at about 7 keV (observer frame).  After
background subtraction PG1244+026 is observed in the GIS up to 10 keV
and NAB0205+024 up to 8 keV (9.3 keV quasar frame), both at the $>3~
\sigma$ level.

Spectral fits were made separately to the spectra from the four ASCA
instruments and to the spectra obtained combining together the data
from the two SIS and GIS detectors. The results were consistent with
each other. In the following we present the results obtained following
the second approach. In some cases $\chi^2$ are smaller than 1. This
is due to the prescription adopted in adding the spectra, for the propagation
of the errors (the Gehrels, 1986, algorithm: error = 1.0 + SQRT(N +
0.75)). Spectra were always rebinned following 2 criteria: a) to
sample the energy resolution of the detectors with four channels at
all energies where possible, and b) to obtain at least 20 counts per
energy channel.  In the spectral fits we limited ourselves to the
0.6-10 keV energy band, to minimize the systematic effect due to the
uncertainty in the SIS calibration below 0.6 keV.  $N_H$ was always
constrained to be greater than or equal to the Galactic value along
the line of sight.

In all cases the quoted errors represent the 90 \% confidence 
interval for 1 interesting parameter.

\section{Variability}

ASCA observed the two quasars for a total elapsed time of
105 ks (PG1244+026) and 119 ks (NAB0205+024). We can therefore study 
the variability of these sources on time scales from a few 
hundred seconds to about 1.5 days.

Figure 1 and 2 show the SIS light curves of the two sources.
Following Ptak et al. (1994) the values were computed using variable
bin sizes corresponding to good time intervals longer than 200
seconds.

\begin{figure}
\epsfig{ file = 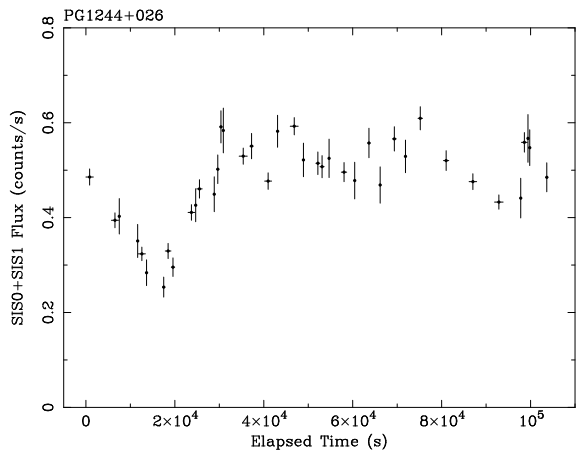, height=8.5cm, angle=-90}
\caption{The SIS 0.5-7 keV light curve of PG1244+026.} 
\end{figure}

\begin{figure}
\epsfig{ file = 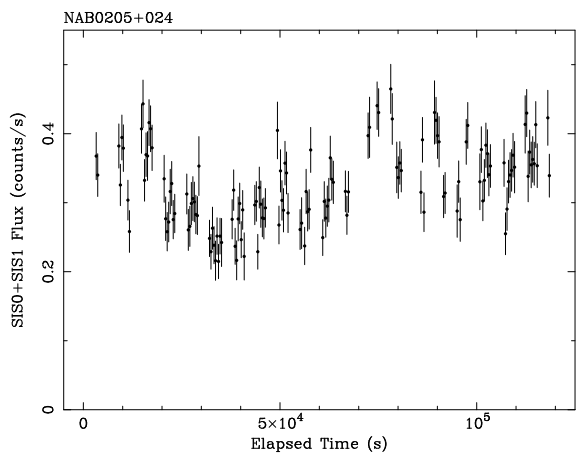, height=8.5cm, angle=-90}
\caption{The SIS 0.5-7 keV light curve of NAB0205+024} 
\end{figure}

Significant variations are evident in both light curves.  A factor of
two variability, both down and up, in about 10 ks is present at the
beginning of the light curve of PG1244+026.  A drop of a factor of
about 2 is present at the beginning of the light curve of NAB0205+024
on a $\sim15$ ks timescale.  In this paper we limit ourselves to
pointing out that roughly similar variability is observed in two
sources which differs in luminosity by a factor $\sim10$.
Significant, although rather small, spectral variability is also
present in the ASCA observations of the two sources.  A systematic
analysis of the variability in different energy bands and of the
spectral variability is in progress and will be presented, together
with a similar analysis on a sample of about 20 Seyfert 1 galaxies and
quasars observed by ASCA and BeppoSAX, in a paper in preparation. In
the following sections we present the average properties of the
spectra. We anticipate that the spectral variability will not modify
the results presented here.

\section{Spectral analysis}

Figures 3 and 4 show the SIS+GIS spectra of PG1244+026 and NAB0205+024
fitted with a simple power law absorbed at low energy by a column of
cold gas equal to or higher than the Galactic column along the line of
sight. Table 3 gives the best fit parameters and the $\chi^2$. It is
clear that this simple model is inadequate to describe the 0.6-10 keV
spectrum of both quasars. A hard tail, larger than the 5-10 \%
systematic uncertainties at these energies (Gendreau \& Yaqoob 1997),
is evident in both cases. In the spectrum of PG1244+026 there is also
a significant excess with respect to the model about 1 keV.  The
feature is visible in both SIS and GIS detectors.  We discuss these
two findings in turn in the next two sections.

\begin{table*}
\centering
\caption{PG1244+026 \& NAB0205+024 spectral fits}
\begin{tabular}{lccccc}
\hline \hline
model    & $N_H^a$  &  $\alpha_E$ or T$^b$ & $\alpha_H$ or $\Omega/2\pi$ 
& $E_{break}^b$ or A$^c$ & $\chi^2$ (dof)\\
\hline
PG1244+026 & && \\
PL 0.6-10 keV    & 1.9+0.6  & 1.67$\pm$0.04 & -- & -- & 177.9 (148) \\
PL 2-10 keV      & 1.9FIXED & 1.35$\pm$0.12 & -- & -- & 48.4 (82) \\
PL 3-10 keV      & 1.9FIXED & 1.03$\pm$0.30 & -- & -- &  29.1 (58) \\
PL 4-10 keV      & 1.9FIXED & 0.67$\pm$0.55 & -- & -- & 14.4 (38) \\
Broken PL 0.6-10 keV & $3.6^{+2.6}_{-1.7}$ & $1.80^{+1.1}_{-0.2}$ 
& $1.06^{+0.24}_{-0.38}$ & 2.8$\pm0.6$ & 149.1 (145) \\
PL+Raym 0.6-10 keV & 1.9+0.7 & $0.88^{+0.2}_{-0.05}$ & 1.54$\pm$0.06 
& $>0.5$ & 135.3 (145) \\ 
PL+BB 0.6-10 keV & $5.3^{+8.0}_{-3.4}$ & 0.16$\pm$0.03 & 1.40$\pm$0.15
& -- & 139.2 (145) \\
PL+Comp.Refl 0.6-10 keV & $2.6^{+1.3}_{-0.7}$ & 2.78$\pm$0.06 & $>8$
& -- & 150.5 (146) \\
\hline
NAB0205+024 & && \\
PL 0.6-10 keV    & 3.0+0.1  & 1.38$\pm$0.02 & -- & -- & 126.2 (147) \\
PL 2-10 keV      & 3.0FIXED & 1.09$\pm$0.10 & -- & -- & 42.4 (86) \\
PL 3-10 keV      & 3.0FIXED & 0.94$\pm$0.20 & -- & -- & 23.3 (58) \\
PL 4-10 keV      & 3.0FIXED & 0.62$\pm$0.45 & -- & -- & 9.5 (38) \\
Broken PL 0.6-10 keV & 3.0+1.4 & $1.50^{+0.3}_{-0.1}$ 
& 0.98$\pm$0.10  & 2.4$\pm0.5$ & 85.2 (145) \\
PL+Raym 0.6-10 keV & 3.0+7.0 & 0.49$\pm$0.05 & 1.02$\pm$0.25 
& $<0.01$ & 82.8 (144) \\ 
PL+BB 0.6-10 keV & 3.0+0.2  & 0.16$\pm$0.02  & 1.15$\pm$0.07 
& -- & 86.2 (145) \\ 
PL+Comp.Refl 0.6-10 keV & 3.0+0.2  & 1.44$\pm$0.04  & $>4.2$ 
& -- & 95.2 (146) \\

\hline
\end{tabular}

$^a$ in $10^{20}$ cm$^{-2}$; $^b$ in keV; $^c$ metal abundances
\end{table*}

\begin{figure}
\epsfig{ file = 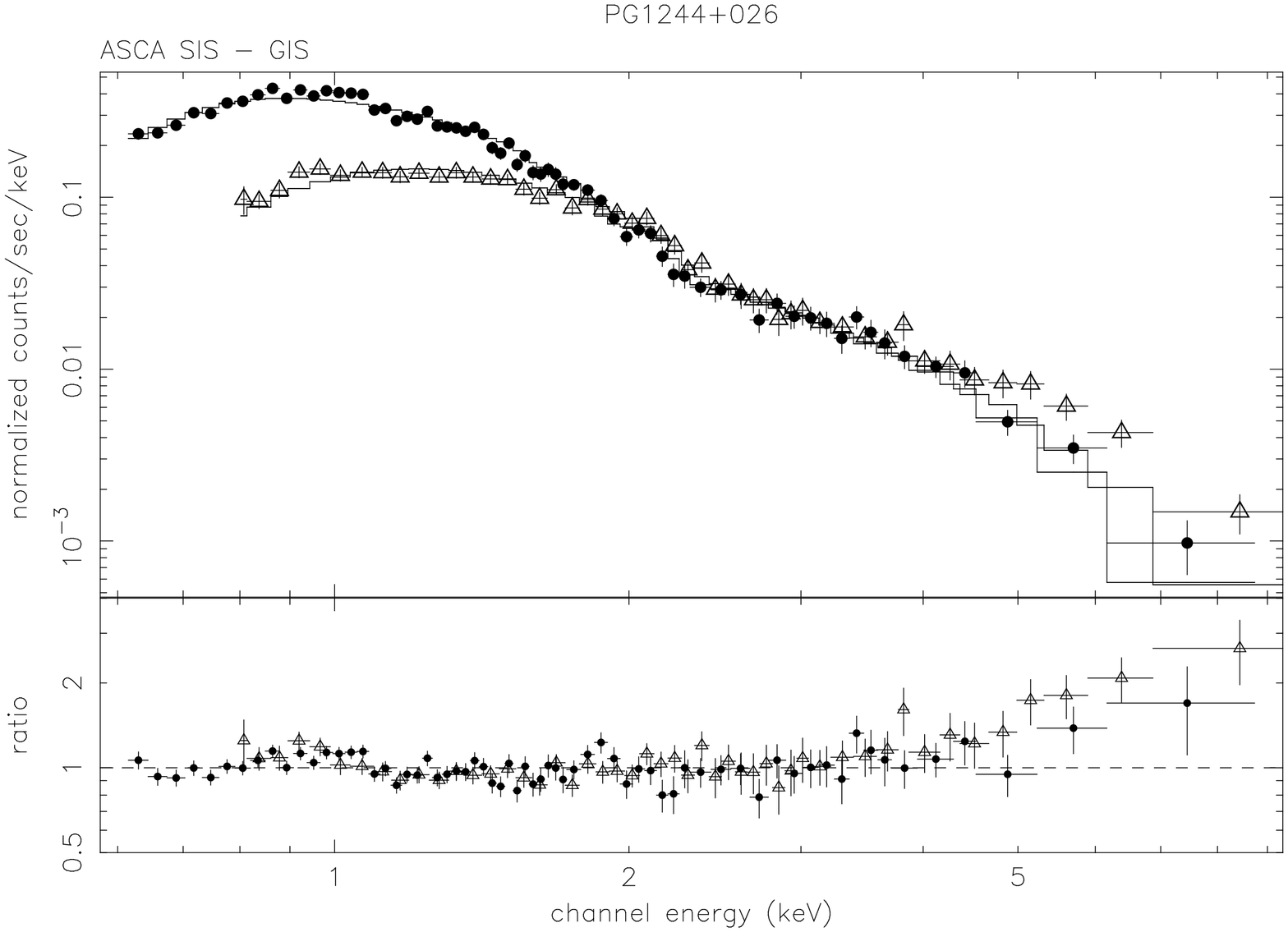, height=9.5cm, angle=-90}
\caption{The SIS+GIS spectra of  PG1244+026 fitted
with a simple absorbed power law model. 
The lower panel shows the ratio between the data and the 
best fit model.}
\end{figure}

\begin{figure}
\epsfig{ file = 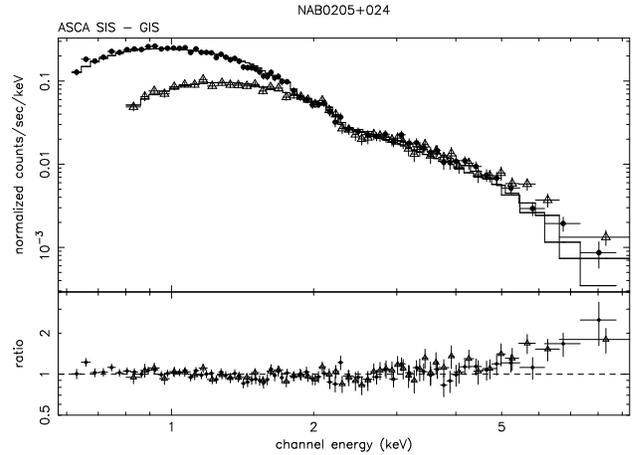, height=9.5cm, angle=-90}
\caption{The SIS+GIS spectra of  NAB0205+024 fitted
with a simple absorbed power law model. 
The lower panel shows the ratio between the data and the 
best fit model.} 

\end{figure}

\subsection {0.6-10 keV continuum}

The ASCA SIS and GIS responses are peaked at 1-2 keV and decrease
sharply at higher energies. Therefore the ASCA `2-10 keV' slopes are
strongly biased toward the lowest energy boundary. This means that
some caution should be used when comparing ASCA `2-10 keV' slopes with
those of experiments whose responses peak around 6~keV such as EXOSAT,
GINGA, BeppoSAX and XTE. To address this, we fitted the SIS+GIS
spectra of PG1244+026 and NAB0205+024 with a simple power law in the
observed 0.6-10, 2-10, 3-10 and 4-10 keV ranges. The results are given
in Table 3 and show in Figure 5.

\begin{figure}
\epsfig{ file = 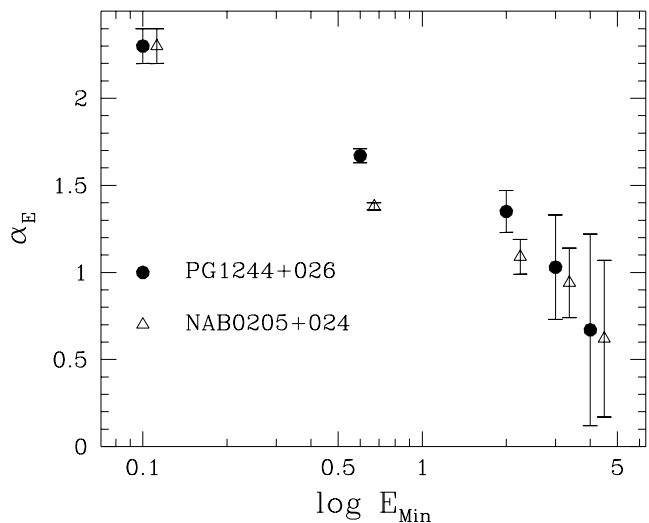, height=9.5cm, angle=0}
\caption{
The PG1244+026 (filled circles) and NAB0205+024 (open triangles)
best fit power law index in the observed
0.6-10, 2-10, 3-10 and 4-10 keV ranges as a function of the 
range minimum energy. The  NAB0205+024 points have been shifted 
from real energy for a sake of clarity. The PSPC 0.1-2 keV points
are also showed for comparison. 
}

\end{figure}
 
In the last three series of fits the $N_H$ was fixed to the Galactic
value. The best fit slopes flatten by 0.5--1 going from the low energy
dominated to the higher energy range. (We note that for NAB0205+024
the quasar redshift implies that these slopes refer to slightly harder
energy ranges: 10 keV in the observer frame corresponds to 11.6 keV in
the quasar frame).

The PSPC spectral index for both quasars is 2.3($\pm$0.1, Fiore et
al., 1994), steeper than any of the values in table 3.  For
NAB0205+024 the ASCA low energy index is flatter than the PSPC by
0.4-0.8, and the ASCA best fit $N_H$ coincides with the Galactic
value, suggesting that the spectrum continues to steepen below ASCA
X-ray energies.

To parameterize a curved spectrum we have also fitted the 0.6-10 keV
spectra of the two quasars with: a broken power law model; a power law
+ optically thin plasma emission (Raymond \& Smith 1977) model; a
power law + black body model; a power law + Compton reflection model.
The results are again in Table 3 and confirm the presence of
significant curvature in the spectra of these quasars.

In PG1244+026 the power law + Raymond-Smith model gives a $\chi^2$
significantly better than the broken power law model because it fits
also the $\sim 1$ keV feature (see next section).  However, the best
fit energy index is still very steep and positive residuals are
evident above 4 keV (Figure 6).  In NAB0205+024 the power law +
Raymond-Smith model gives an acceptable $\chi^2$ only for low metal
abundances, given the absence of significant features in the spectrum,
similar to the ROSAT results of Fiore et al. (1994). Above 4 keV the
best fit is similar to that in the broken power law model.  Fits with
a power law + black body models give acceptable $\chi^2$ in both
cases. The power law indices in this case are slightly steeper than in
the broken power law model. Inspection of the residuals shows again a
slight excess of counts at high energy. In PG1244+026 the power law +
black body model again gives a $\chi^2$ significantly better than the
broken power law model because it partly fits also the feature around
$\sim 1$ keV (see next section).  Fits with a Compton reflection model
({\sc plrefl} in {\sc xspec}) give $\chi^2$ significantly higher than
the previous models and push the parameter $\Omega/2\pi$ to high and
implausible values.

\begin{figure}
\epsfig{ file = 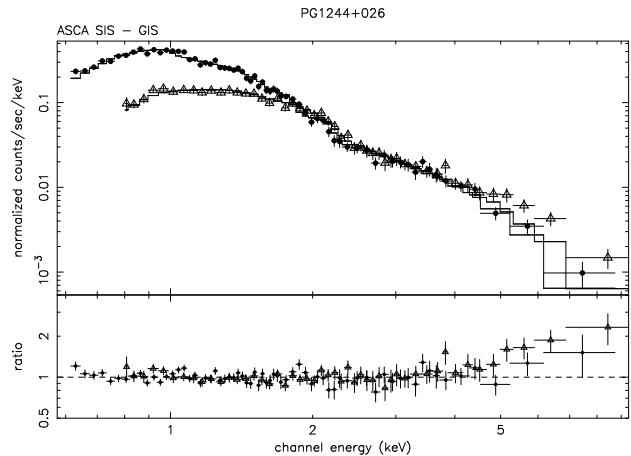, height=9.5cm, angle=-90}
\caption{The SIS+GIS spectra of  PG1244+026 fitted with a 
power law + Raymond-Smith model.
The lower panel shows the ratio between the data and the model} 
\end{figure}

We do not see any significant line emission at the energies of the
iron K$\alpha$ lines. The 90 \% upper limits to the equivalent width
of a narrow line at 6.4 (and 6.7 keV), rest frame, in  PG1244+026 and
NAB0205+024 are 400 eV and 230 eV (640 eV, 314 eV) respectively.

The $\chi^2$ for the fits to the PG1244+026 spectra are much higher than
those for the similar fits to the NAB0205+024 spectra because of the
presence of the `1 keV' feature in the former quasar. We discuss
this feature next.

\subsection {Low energy features}

We have already noted that a `1~keV' feature seems to be present only
in PG1244+026, the quasar with the lower luminosity.  Figure 7 shows
the results of a simple power law model fit to the 0.6-4 keV spectrum
of PG1244+026: the `1 keV' feature is clearly visible. 

\begin{figure}
\epsfig{file=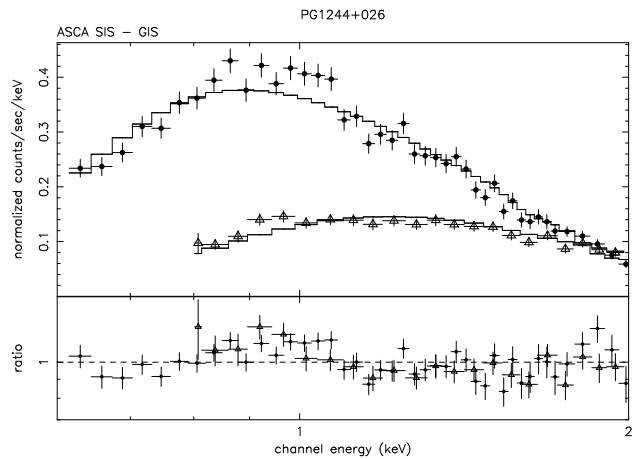, height=9.5cm, angle=-90}
\caption{The SIS+GIS spectra of PG1244+026 fitted in the 0.6-4 keV range.
The lower panel shows the residuals
when the spectrum is fitted with a simple absorbed power law} 
\end{figure}

It has been recently realized that Residual Dark Distribution (RDD,
the error in ASCA's onboard correction for CCD dark current) can
affect the low energy ASCA SIS spectra. The symptom of the RDD problem
is a sudden decrease in the SIS effective area towards lower energies
from about 1 keV.  We are however confident that this has little
effect on the `1 keV' feature for the following reasons: a) SIS data
of PG1244+026 and NAB0205+024 have similar RDD value but the `1 keV'
feature is visible in the spectrum of PG1244+026 only, the 90 \% upper
limit to any unresolved gaussian line emission at about 1 keV in
NAB0205+024 is of 20 eV; b) the feature is present also in the GIS
spectrum; the feature is present at the same level in SIS0 and SIS1,
while the RDD tends to be greater for SIS1 than for SIS0.

To investigate the physical origin of the `1 keV' feature and to
quantify its strength we performed a series of spectral fits to the
SIS and GIS spectra of PG1244+026 using only the 0.6-4 keV range, to
avoid the complications of the hard tail (see previous section). The
results are reported in Table 4, along with those of the other fits
in this section.

\vskip .25in
\begin{table*}
\centering
\caption{PG1244+026: spectral fits in the band 0.6-4 keV}
\begin{tabular}{lcccccc}
\hline \hline

\multicolumn{7}{l}{Continuum models}\\
model    & $\alpha_E$ & $N_H^a$   & T$^b$ & A$^c$ &  &  $\chi^2$ (dof) \\
\hline

PL       & 1.70$\pm$0.02 & 1.9+1.0 & -- & --  & -- & 131.8 (104) \\

PL + BB  & 1.50$\pm$0.11 & 9.5$\pm$5.0 & $0.14\pm0.02$ & -- 
& -- & 109.5 (101) \\

PL + Raym& 1.61$\pm$0.03 & 1.9+0.3 & 1.00$\pm$0.06 & $>$0.5 
& -- & 93.1 (100) \\
\hline

\multicolumn{7}{l}{emission line} \\
        &      &      & line E$^b$ & line EW$^d$ &  & $\chi^2$ (dof) \\
\hline

PL + Gauss & 1.60$\pm$0.04 & 1.9+0.5 & 0.91$\pm$0.03 & 64$\pm$15 
& -- & 90.1 (101) \\
\hline

\multicolumn{7}{l}{absorption edge models}\\
        &    &   & edge E$^b$ & $\tau$ &     & $\chi^2$ (dof) \\
\hline

PL + 1 edge  & 1.88$\pm$0.11 & 8.2$\pm$3.0 & 1.12$\pm$0.03 & 0.25$\pm$0.07 
& -- & 102.5 (102) \\

PL + 2 edges & 1.88$\pm$0.11 & 1.9FIXED & 1.12$\pm$0.10 & 0.26$\pm$0.08 
& -- & 94.8 (101)\\
          &               &          & $<0.64$ & 0.51$\pm$0.14  
& \\
\hline

\multicolumn{7}{l}{ionized absorber models}\\
       & & & log$N_H$(ionized) & U or logT & 
z abs. & $\chi^2$ (dof) \\
\hline
1 -- Phot. ion.  & 1.92$\pm$0.10 & 11.1$\pm$3.5 & 22.59$\pm$0.15 
& 2.00$\pm$0.12 & 0.25$\pm$0.07 & 93.2 (101) \\
2 -- Phot. ion.  & 1.94$\pm$0.10 & 11.0$\pm$3.5 & 21.56$\pm$0.10 
& 1.20$\pm$0.10 & -0.33$\pm$0.10 & 95.7 (101) \\
3 -- Coll. ion.  & 1.90$\pm$0.10 & 11.3$\pm$3.5 & 22.62$\pm$0.14
& 6.92$\pm$0.07 & 0.24$\pm$0.06 & 90.7 (101) \\
\hline

\multicolumn{7}{l}{absorption notch models}\\
       &  &   & T$^b$, notch E$^b$ & notch width$^d$ & Cov. frac. 
& $\chi^2$ (dof) \\
\hline

PL + notch & 1.93$\pm$0.08 & 9.3$\pm$3.3 & 1.44$\pm$0.05 & 680$\pm$70 
& 0.14$\pm$0.04 & 97.5 (100) \\
PL+BB+notch & 1.48$\pm$0.20 & 7.8$\pm$4.4 & $T=0.16\pm0.03$  &  
&               & 97.1 (100) \\
            &               &             & 1.16$\pm$0.03 & 14$\pm$7 
& 1FIXED & \\

\hline
\end{tabular}

$^a$ in $10^{20}$ cm$^{-2}$; $^b$ in keV; $^c$ metal abundances; 
$^d$ in eV
\end{table*}

\subsubsection{Emission Line model}

The fit with a power law model plus a gaussian line gives a small
$\chi^2$ (90.1 for 101 degrees of freedom).  The equivalent width of
the 1 keV feature, in the power law plus gaussian line fit, is similar
in the SIS and GIS detectors. In Table 4 we report the SIS
determination of 64$\pm$15 eV.  The line width is well constrained and
cleanly resolved in the SIS spectrum to $\sigma=0.11^{+0.01}_{-0.03}$
keV.

A fit with a power law plus a thermal plasma model (Raymond-Smith
1977) gives an acceptable $\chi^2$ (93.1, 100 dof). The best fit
temperature (1 keV) implies that the `1 keV' emission is dominated by a
blend of iron-L and neon emission lines.

An emission line feature can be mimicked by fitting a spectrum with a
strong absorption feature at slightly higher energies. We have then
fitted the SIS and GIS spectra with models including absorption
structures.

\subsubsection{Absorption Edge Fits}

Fits with one absorption edges gives $\chi^2$ significantly higher
than the previous case (102.5, 102 dof).  The edge energy (1.17 keV,
quasar frame) is consistent with that of NeIX and/or iron L FeXVI and
FeXVII.  The best fit neutral $N_H$ is significantly higher than the
Galactic value.  This is reasonable, since if there is highly ionized
Ne and Fe L absorption it is likely to have also highly ionized oxygen
absorption at 0.74-0.87 keV.  We then refitted the SIS and GIS spectra
with a model including three absorption edges, at the energies of the
most abundant ions in highly ionized gas with high NeIX abundance:
OVII, OVIII and NeIX-FeXVI-FeXVII (Nicastro et al. 1998), fixing the
cold $N_H$ to the galactic value.  The results were not
satisfactory. The depth of the oxygen edges is zero with small upper
limits and the $\chi^2$ is significantly higher than in the previous
case: 114.7.  Leaving $N_H$ free improves the $\chi^2$, but the oxygen
edge depths are still zero and the fit resembles completely the single
edge fit.  Fixing $N_H$ to the Galactic value but leaving free the
energies of two edges produces again a good fit ($\chi2=94.8$, 101
dof).  The best fit energy of one edge is again 1.12 keV, but that of
the other edge is $<0.64$ keV (observer frame), close to the lower
boundary of the observed range. So, if the cold absorption is fixed to
the Galactic value, then there must be additional absorption edge(s)
at energies lower than the observed range, corresponding to oxygen
less ionized than OVI.  We note however that this conclusion is
weakened by the unknown contribution of the SIS RDD, which
pushes low energy events below the detection threshold.

A 1.17 keV absorption feature can also be interpreted in terms of
blueshifted oxygen absorption (Leighly et al. 1997).  In this case,
assuming that the absorption is mostly due to OVIII, the shift from
the quasar frame would be equivalent to z=--0.38.  A more complex
continuum has little effects on the best fit parameters of absorption
edges.

\subsubsection {Ionized absorber models Fits}

We fitted the data with a detailed ionized absorber model (not
including resonant scattering absorption lines).  We first generated a
grid of photoionization equilibrium models using {\sc Cloudy} (Ferland
1996), and fitted the spectrum interpolating by this grid, using the
method of Fiore et al. (1993). To calculate the models we have assumed
the observed spectral energy distribution (Fiore et al. 1995, Elvis et
al. 1994). This is important, since the soft X-ray spectrum of this
source strongly differs from that of `normal' Seyfert 1 galaxies,
where warm absorbers are usually found (Reynolds 1997). A steep soft
X-ray spectrum can completely ionize oxygen and neon but not iron, and
so can produce edges in the 1-2 keV (Fe-L) and 7-9 keV (Fe-K) ranges,
but not in the `oxygen' 0.6-0.9 keV band.  A fit with this model
produces an acceptable $\chi^2$ (see Table 4, ionized absorber model 1).

In Figure 8 we show the best fit steep SED model (thick line), and a
photoionization model obtained using a standard, much flatter AGN SED
(a power law of $\alpha=1.2$ from UV to X-rays, thin line), which,
above 1 keV, gives a comparably good fit to the data. While iron in
the flat SED model has a ionization structure similar to that of the
model obtained using the right SED, oxygen is much less ionized: note
the deep OVIII edge present in the flat SED model.

In this fit the redshift of the absorber is significantly higher than
that of the quasar, because the main feature in the transmitted
spectrum is the FeXVIII edge at 1.36 keV, while the deepest edge in
the quasar spectrum is at 1.17 keV (quasar frame).  However, a good
fit can be also obtained for a different absorber redshift (z=--0.33,
Table 4, ionized absorber model 2). In this case the 1.17 feature is
interpreted in terms of OVII and OVIII absorption.  We cannot
discriminate between these two solutions on statistical grounds.

We have also tried fits with a collisional equilibrium model (Nicastro
et al. 1998). The results were very similar to those obtained in the
case of photoionization equilibrium (see Table 4, ionized absorber
model 3).

In all fits with detailed warm absorber models the column of cold gas
is significantly higher than the Galactic, value, similar to the
values found using single edges to parameterize the absorber (Table
4).

\begin{figure}
\epsfig{file=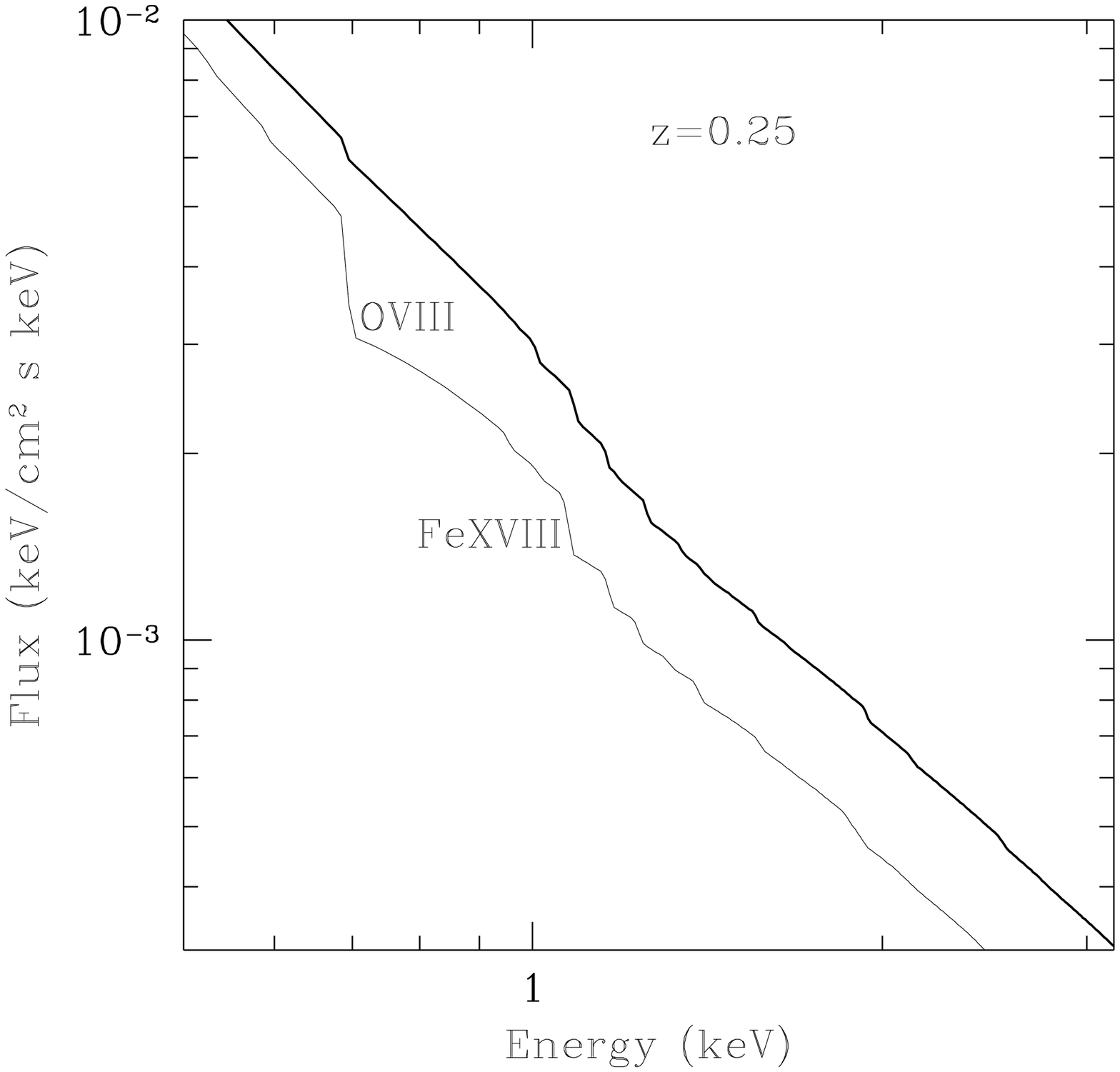, height=9.5cm }
\caption{The thick line shows the best fit photoionized absorber model 
for PG1244+026 using the observed steep spectrum
(note that the absorber has z=0.25, while the
quasar has z=0.048). The thin line shows a photoionization model 
obtained using a much flatter SED (see text). Both models produce Fe-L
absorption features around 1 keV. The flat SED also produces
deep OVIII edge, which is not observed} 
\end{figure}

%

\subsubsection{Absorption Line Fits}

Fits with a single gaussian absorption line do not give acceptable
$\chi^2$. Fits with two or more absorption gaussian lines can produce
$\chi^2$ of 96 or smaller. These models are indistinguishable for the
SIS from models with a broad absorption notch, which we discuss in the
following. Physical models including resonant absorption lines, as
well as absorption edges from ionized plasma will be discussed in a
paper in preparation (Nicastro et al. 1998b).

\subsubsection{Absorption Notch Fits}

Fits with an absorption notch give $\chi^2$ 
higher than those with an emission lines by $\Delta\chi^2\approx 7$.
While the power law + notch fit is formally acceptable, the best fit
value of the notch width is implausibly large (almost 1 keV), forced
by the very low value of the covering fraction required by the fit.
However, the notch best fit parameters are strongly dependent on the
proper modeling of the continuum. For example, using a power law + a
black body to parameterize the continuum gives an acceptable fit
fixing the notch covering fraction to 1, which in turn results in a
much more reasonable value for the notch width of 14$\pm$7 eV. The
ASCA band width in not wide enough and its spectral resolution is not
good enough to constrain adequately {\it both} a complex continuum
{\it and} the notch parameters.

\subsubsection{Comparison with the PSPC results}

PG1244+026 was observed with the PSPC in December 1991 and the
results of this observation have been reported by Fiore et al. (1994).
The 0.6-2 keV flux level during the PSPC observation was about 50 \%
lower than the mean flux in the ASCA observation. The fit of an
absorbed power law to the PSPC spectrum gives an acceptable $\chi^2$
(22.8 for 25 dof), $\alpha_{0.1-2keV}=2.3\pm0.1$ and
$N_H=2.9\pm 0.3\times10^{20}$ cm$^{-2}$, slightly higher than the
Galactic value. Most of the PSPC counts were detected below 0.3 keV,
in the `Carbon' band. The quality of the spectrum between 0.3 and 2
keV is not very high and emission or absorption features fainter
than $\sim 20 \%$ cannot be excluded in this energy band. No evidence
of spectral variability is present in the ROSAT data despite a factor
of 2 flux variability.

The PSPC data strongly constrain the level of any cold or warm
absorption affecting the `Carbon' band. Best fitting models to the
ASCA data including absorption features in the 1-2 keV band (Table 4)
require a rather large absorption in addition to the Galactic one
below 1 keV.  Therefore, it is important to study whether the ASCA
best fit models are consistent with the PSPC ones.  Rather than
performing joint fits to the ASCA and PSPC data, which are complicated
by the large uncertainty in the relative PSPC/ASCA SIS calibration,
and by the detailed shape of the continuum over the broad 0.1-4 keV
band, we fitted the PSPC data with a power law model including the
emission or absorption features found in the previous section (see
Table 4).  The results are in Table 5, where we also report (in
brackets) the 99 \% parameter upper limits, or confidence intervals,
when appropriate.

\begin{table*}
\centering
\caption{PG1244+026: spectral fits to the PSPC}
\begin{tabular}{lccccc}
\hline \hline
model    & $N_H^a$   & $\alpha_E$  &  line or edge E$^b$ & 
line EW$^c$ or $\tau$ & $\chi^2$ (dof) \\
\hline
PL           & 2.9$\pm$0.3 & 2.3$\pm$0.1 &  & & 22.8 (25) \\
PL + gauss   & 2.9$\pm$0.3 & 2.3$\pm$0.1 & 0.91FIXED & 0+68 (+100)
& 22.8 (23) \\
PL + 1 edge  & 2.7$\pm$0.4(0.6) & 2.2$\pm$0.2 & 1.12FIXED & $0.3^{+0.4}_{-0.3}$
& 20.9 (24) \\
PL + 2 edges & 2.7$\pm$0.4      & 2.2$\pm$0.2 & 1.12FIXED & $0.3^{+0.4}_{-0.3}$
& 20.9 (23) \\
             &                &           & 0.62FIXED & 0+0.34(+0.62)
& \\ 
\hline
\end{tabular}

$^a$ in $10^{20}$ cm$^{-2}$; $^b$ in keV; $^c$ in eV
\end{table*}

We see that the presence of an emission line at 0.91 keV is not
required by the PSPC spectrum, but an equivalent width of 65 eV is not
excluded (10 \% probability). However, the presence of a cold absorber
of thickness $8.2\pm3.0\times10^{20}$ cm$^{-2}$ is inconsistent with
the PSPC spectrum (probability $<1\%$), while the presence of an edge
at 0.62 keV with $\tau=0.51\pm0.14$ is only marginally consistent with
the PSPC result.

\section{Discussion}

\subsection{Continuum}

The ASCA observations of PG1244+026 and NAB0205+024 have shown that
the X-ray continuum of these two quasars flattens by 0.5-1 passing
from the 0.1-2 keV (PSPC) to the 2-10 keV band. Similar results were
obtained by Brandt, Mathur \& Elvis (1997); and by Comastri et
al. (1998) and Leighly et al. (1998a) on TONS180, Pounds et al. (1995)
and Fiore et al (1998b) on REJ1034+390, Leighly et al. (1998b) on
AKN564.  It appears that the X-ray spectrum of a sizeable number of
steep PSPC and narrow Balmer line quasars has significant curvature,
being flatter at higher energies.

This could be due to different components influencing the spectrum at
different energies, as might happen in `normal' broad lines Seyfert 1
galaxies and quasars, where a soft component is often present.  The
relative intensity of the two components would be quite different from
`normal' quasars.  Laor et al. (1997) suggested a fainter hard
component relative to the optical in the majority of low redshift PG
quasars (assuming a two component model). However, Grupe (1996) found
evidence for a stronger soft excess in a sample of soft X-ray selected
Seyferts dominated by narrow-line objects.

A large relative intensity of the soft component has important
consequences on various competing models for the soft component.  Disc
reprocessing models (Matt et al. 1993, Fiore et al. 1997) would
require highly anisotropic emission to account for the discrepancy
between the observed soft and hard fluxes.  Optically thin free-free
emission (e.g. Barvainis 1993) is ruled out by these observations
because the best fit power law slope is still too steep to fit the
spectrum above 4 keV, because variability rules out optically thin
plasma (Elvis et al. 1991) and because of the implausibly low metal
abundances ($<1 \%$ solar, see Table 3) required in NAB0205+024 (see
Sect. 3.2.1).  The most likely origin for the steep component is
Comptonized disc emission (e.g. Czerny \& Elvis 1987, Fiore et
al. 1995, Pounds et al. 1995).

The high energy spectral index of the two quasars $\alpha_H\sim1.0$
(see Table 3) is consistent with that of `normal' Seyfert 1 galaxies
(e.g. Nandra \& Pounds 1994, Nandra et al. 1997). The error on
$\alpha_H$ is however large and so no strong conclusion can be drawn on
the origin of the hard emission.
An answer to this question must await
the large area and high energy sensitivity of AXAF,
XMM and Spectrum X-gamma.

\subsection{Origin of the $\sim1$ keV feature in PG1244+026}

The $\sim1$ keV feature in PG1244+026 could be explained in terms of
either a broad ($\sigma=0.1$ keV) emission line at 0.91 keV (0.95 keV
quasar frame) of about 60 eV equivalent width or a $\tau=0.25$
absorption edge at 1.17 keV (or an absorption notch at 1.22 keV).  
These possibilities cannot be discriminated between on
statistical grounds.

An absorption interpretation requires additional low energy
absorption, either cold (with a column density higher than Galactic by
$\sim7\times10^{20}$ cm$^{-2}$ (see Tables 4 and 5), in contrast with
the PSPC results, or a warm absorber with a peculiar ion abundance
distribution.

The observed spectrum can be interpreted in terms of either an inflowing
(v/c=0.25) or an outflowing (v/c=--0.33) absorber.  In the first case the
ion contributing most to the absorption is FeXVIII, in the second case
it is OVIII.  The two cases cannot be discriminated on statistical
grounds. While an outflowing absorber has been
suggested in several other cases (see e.g. Mathur et al. 1994), this
would be the first case for an inflowing highly ionized absorber.
 
A similar situation is found in IRAS 13224-3809 by Otani et
al. (1995), in AKN564 by Brandt et al.  (1994) and by Leighly et
al. (1998a).  Otani et al (1995) and Leighly et al. (1997) interpret
the features in the 1--2 keV band in terms of blueshifted absorption
from relativistically (v=0.2-0.6 c) outflowing material.  If the 1.17
keV feature in PG1244+026 is due to a blueshifted OVIII photoelectric
absorption, then the absorption seen below 0.64 keV may be due to CVI
photoelectric absorption from the same gas. This distribution is far
from an equilibrium distribution (see eg. Nicastro et al. 1998), not
an impossible situation considering the large variability observed in
this source.

The absorption features seen between 1 and 2 keV may also be
interpreted in terms of resonant lines (e.g. Leighly et al. 1997). If
the ion producing the absorption is oxygen OVIII (resonant line at
0.65 keV), the best fit notch energy of 1.22 keV (quasar frame)
implies a very high gas velocity: --0.56c.  Fe XVIII (E=0.87 keV) or
Fe XVII (E=0.81 keV) can also contribute to the absorption, because of
their high oscillator strengths, 1.7, 0.6 respectively (Kato et
al. 1976), and abundances. If the 1.22 keV absorption notch is due to
these ions then the velocity of the outflowing gas will be smaller,
--0.33c.

In any case, in the blueshifted absorption scenario the gas is
outflowing at velocities which are a sizeable fraction of c,
reminiscent of blobs of gas in jets. It is interesting to note that
similar absorption features have been observed in Blazars
(e.g. PKS2155-304, Canizares \& Kruper, 1884, other BlLacs, Madejski
et al. 1991, 3C273, Grandi et al. 1997), but usually below 1 keV.
Somewhat surprisingly this implies less extreme conditions in these
radio-loud objects than in our radio-quiet quasars.  High redshift
radio loud quasars may have similar jet-related absorption too (Elvis
et al., 1997).

An alternative interpretation of the 1 keV feature is in terms of an
emission line due to highly ionized oxygen, neon (NeIX) and/or to Iron
L. There are two possible origin for this line: recombination in an
optically thin thermal plasma, or ``reflection" in photoionized
matter. 

\subsubsection{Thermal plasma}

A thermal plasma is highly implausible on physical grounds (Elvis et
al. 1991, Fiore et al. 1995).  Emission measure is
$\sim2.5\times$10$^{65}$ cm$^{-3}$, For a spherical source (with
radius $R$) and constant electron density $n_e$, $n_e=2.43\times
10^{32} R^{-{3 \over 2} }$ cm$^{-3}$.  $R<10^{14}$ cm from the observed
X-ray variability, $n_e<1.5\times$10$^{12}$ cm$^{-3}$, This implies an
electron scattering optical depth $\tau_{\rm T} \gs 30$.  With such
values a thermal plasma is no longer optically thin. The situation is
even worse if the matter is clumpy, as the density of each cloud must
be greater.

\subsubsection{Photoionized matter}

The second possibility is that the `1 keV' emitting matter is
photoionized by the central nucleus. We can assume that the gas is not
covering the source because there is not significant OVII and/or OVIII
absorption.  We therefore assume that we are not observing a ``warm
absorber", but rather a ``warm reflector''. This could either be a
warm absorber viewed from its side, in which case the matter would be
optically thin to Thomson scattering; or the accretion disc, and the
matter would be thick. Since both the ``reflector'' and the primary
emission (which provides most of the continuum) are observed, the
optically thick case gives the highest values of the equivalent
width. However, even in the optically thick case the expected EWs can
barely account for the observed values (see below); hence we neglect
the optically thin case altogether.

The observed line (which is significantly broad) may be a blend of
several lines (see for instance \.Zycki et al. 1994, Netzer 1997).  The
observed energy suggests the 0.92 keV Ne {\sc ix} and 1.02 keV Ne {\sc
x} recombination lines, the 0.87 keV Oxygen {\sc viii} recombination
(to ground state) line and the iron L (around 0.8 keV) lines being the
most important. None of these lines alone can account for the observed
EW: for instance, the maximum value (i.e. for a face--on disk with a
intervening ion fraction $\sim0.6$) for the Ne {\sc
ix} line is about 10 eV, while that of the oxygen recombination line
is about 15 eV (note that a O {\sc viii} K$\alpha$ recombination line
at 0.65 keV with a similar EW should also be present; the 90\% upper
limit on such a line is 30 eV).  These values have been calculated
using the formulae of Basko (1978), and assuming a reasonable
ionization structure. (In the disc hypothesis an iron
K$\alpha$ line at 6.5-6.9 keV is also expected, but the upper limit of 
300-400 eV does not exclude the presence of such a line.)
Allowing for a possible factor of 2 neon and/or iron
overabundance (an oxygen overabundance would decrease the Ne line while
not increasing the O line) and/or anisotropy of the illuminating
radiation, the observed equivalent width could be explained (note that
in many Seyfert 1 galaxies the iron K$\alpha$ line is also stronger
than expected, suggesting iron overabundance or anisotropic
illumination).

In this scenario the `1 keV' feature may arise from an highly ionized
accretion disk. In the Matt et al. (1993) models high ionization is
mainly due to a high accretion rates (the ionization parameter depends
on $\dot m^3$).  The detection of these emission lines in steep soft
X-ray quasars would then be further evidence of high $L/L_{\rm Edd}$.
We note that since recombination can occur only in highly ionized
atoms, we would not expect features of this kind in `normal' quasars,
as their disc should be much less ionized, as in fact observed.

The strong dependence of the ionization parameter on $\dot m$ allows
for large differences in the ionization structure against small
differences in $\dot m$ and therefore that the `1 keV' feature may not
be ubiquitous in NLSy1s.  Indeed, a similar feature is not present in
NAB0205+024. A line with the same equivalent width as in PG1244+026
(60 eV) would have been detected in the NAB0205+024 SIS spectrum (the
90 \% upper limit is only 20 eV).  In the accretion disc scenario this
would imply a different ionization state (higher or lower) of the
matter or a significant metal underabundance in NAB0205+024.

\subsection{Variability}

The mean 0.5-10 keV luminosity measured by ASCA in PG1244+026 and
NAB0205+024 differs by an order of magnitude: $7.3\times 10^{43}$ and
$6.4\times10^{44}$ erg s$^{-1}$ respectively (assuming isotropic
emission, $H_0$ = 50.0 and $q_0$ = 0.0). The optical (3000 \AA)
monochromatic luminosities differs even more: $4\times10^{43}$ and
$9.4\times10^{44}$ erg s$^{-1}$ respectively. The variations seen in
the NAB0205+024 light curve imply an efficiency in the conversion of
matter into radiation greater than 1.6 \% (e.g. Fabian 1984). Taken at
face value, this excludes thermonuclear reactions as the origin of the
observed X-ray luminosity for which the upper limit on the efficiency
is in this case about 0.7 \%.

The similar variability observed in the two quasars agrees with the
Fiore et al. (1998a) finding that the variability properties of (PG)
quasars are correlated with the shape of the soft X-ray spectrum and
the width of the Balmer lines, and so possibly then with the accretion
rate, in the scheme of Pounds et al. (1995), and Laor et al. (1994),
(1997).

\section{Conclusions}

ASCA observations of two steep soft X-ray quasars have
shown that:

\begin {enumerate}

\item
The X-ray continuum of the two quasars flattens by
$\Delta\alpha=0.5-1$ going toward high energies. Similar results were
obtained by authors on some half dozen steep soft X-ray quasars.

\item
PG1244+026 shows a significant feature in the `1 keV' region.  Similar
features were again reported in other steep soft X-ray quasars.
The data are not good enough
to discriminate between a broad emission line centered at 0.95 keV
(quasar frame) or an absorption edge at 1.17 keV, or
an absorption notch at 1.22 keV.
  
Line emission could be due to reflection from an highly ionized
accretion disk, in line with the view that steep soft X-ray quasars
are emitting close to the Eddington luminosity.  Photoelectric edge
absorption or resonant line absorption could be produced by gas
outflowing at a large velocity (0.3-0.6 c). In these absorption models
significant cold (i.e. oxygen less ionized than OVI) absorption in
excess of the Galactic is required. This would imply an increase by a
factor 2-3 of the cold column with respect to a previous PSPC
observation or a peculiar ionization structure.  In neither the
emission or absorption cases the SIS resolution is good enough to
identify unambiguously the ions responsible for the feature. The high
resolution and high throughput of the low energy gratings and
spectrometers of AXAF and XMM are clearly needed to shed light on this
puzzling case.

\item
The two quasars show similar variability properties (flux variations
up to a factor of 2 in 10 ks) despite a factor of ten difference in
the X-ray observed luminosity. This agrees with the Fiore et al
(1998a) finding that the variability properties of radio-quiet quasars
are correlated with the shape of the X-ray spectrum, the width of the
Balmer lines and so possibly with the accretion rate.

\end{enumerate}

\bigskip
F.F acknowledges support from NASA grants NAG 5-2476 and NAG 5-3039,
B.J.W. acknowledges support from ASC contract NAS8-39073.

\end{document}